\documentclass[aps,twocolumn,showpacs,superscriptaddress]{revtex4}
\usepackage{epsfig}
\usepackage{times}

\begin{document}

\title{Evolutionary mixed games in structured populations: Cooperation and the benefits of heterogeneity}

\author{Marco A. Amaral}
\email{marcoantonio.amaral@gmail.com}
\affiliation{Departamento de F\'\i sica, Universidade Federal de Minas Gerais, Caixa Postal 702, CEP 30161-970, Belo Horizonte - MG, Brazil}

\author{Lucas Wardil}
\affiliation{Departamento de Fisica, Universidade Federal de Ouro Preto, Ouro Preto, MG, Brazil}

\author{Matja{\v z} Perc}
\affiliation{Faculty of Natural Sciences and Mathematics, University of Maribor, Koro{\v s}ka cesta 160, SI-2000 Maribor, Slovenia}
\affiliation{CAMTP -- Center for Applied Mathematics and Theoretical Physics, University of Maribor, Krekova 2, SI-2000 Maribor, Slovenia}

\author{Jafferson K. L. da Silva}
\affiliation{Departamento de F\'\i sica, Universidade Federal de Minas Gerais, Caixa Postal 702, CEP 30161-970, Belo Horizonte - MG, Brazil}

\begin{abstract}
Evolutionary games on networks traditionally involve the same game at each interaction. Here we depart from this assumption by considering mixed games, where the game played at each interaction is drawn uniformly at random from a set of two different games. While in well-mixed populations the random mixture of the two games is always equivalent to the average single game, in structured populations this is not always the case. We show that the outcome is in fact strongly dependent on the distance of separation of the two games in the parameter space. Effectively, this distance introduces payoff heterogeneity, and the average game is returned only if the heterogeneity is small. For higher levels of heterogeneity the distance to the average game grows, which often involves the promotion of cooperation. The presented results support preceding research that highlights the favorable role of heterogeneity regardless of its origin, and they also emphasize the importance of the population structure in amplifying facilitators of cooperation.
\end{abstract}

\pacs{89.75.Fb, 87.23.Ge, 89.65.-s}
\maketitle

\section{Introduction}
\label{Introduction}
Evolutionary game theory \cite{maynard_82, weibull_95, hofbauer_98, mestertong_01, nowak_06} is a powerful theoretical framework for studying the emergence of cooperation in competitive settings. The concept of a social dilemma is particularly important, where what is best for an individual is at odds with what is best for the society as a whole. Probably the most often studied social dilemma is the prisoner's dilemma game \cite{axelrod_84}. During a pairwise interaction, each player can choose either to cooperate or to defect. If both players choose to cooperate they receive the reward $R$, while mutual defection leaves both with the punishment $P$. A defector exploiting a cooperator receives the highest payoff, the temptation $T$, whereas the exploited cooperator receives the sucker payoff $S$. The typical payoff ranking for the prisoner's dilemma game is $T>R>P>S$ \cite{nowak_06, szabo_pr07, perc_bs10}. Evidently, whichever strategy the opponent chooses, it is always better to defect. If both players are rational and adhere to this, they both end up with a payoff that is lower than the one they would obtained if they had chosen to cooperate. Despite its simplicity, however, the iterated prisoner's dilemma game continues to inspire research across the social and natural sciences \cite{santos_prl05, imhof_pnas05, wu_zx_pa07, gomez-gardenes_prl07, tanimoto_pre07, poncela_njp07, du_wb_pa08, du_wb_pa09c, poncela_epl09, tanimoto_pre12, arenas_jtb11, gomez-gardenes_srep12, requejo_pre12b, szolnoki_pre14, tanimoto_pre13, tanimoto_amc15, matamalas2015strategical, javarone_csn15, javarone2016statistical}. If the ranking of the payoffs is changed, other social dilemmas, such as the snowdrift game for $T>R>S>P$, are obtained, which has also received substantial attention in the recent past \cite{sysiaho_epjb05, wang_wx_pre06, du_wb_epl09, chen_xj_epl10, laird_ijbc12, laird_pre13, jiang_ll_pone13}.

Although defection is the rational choice, cooperation in nature abounds. Eusocial insects like ants and bees are famous for their large-scale cooperative behavior \cite{wilson_71}, breeding in birds prompts allomaternal behavior where helpers take care for the offspring of others \cite{skutch_co61}, and chief among all, we humans have recently been dubbed supercooperators \cite{nowak_11} for our unparalleled other-regarding abilities and our cooperative drive. This fact constitutes an important challenge to Darwin's theory of evolution and natural selection, and accordingly, ample research has been devoted to the identification of mechanisms that may lead to cooperative resolutions of social dilemmas. Classic examples reviewed in \cite{nowak_s06} include kin selection \cite{hamilton_wd_jtb64a}, direct and indirect reciprocity \cite{trivers_qrb71, axelrod_s81}, network reciprocity \cite{nowak_n92b}, as well as group selection \cite{wilson_ds_an77}. Diffusion and mobility have also been studied prominently \cite{vainstein_pre01, vainstein_jtb07, wang_z_srep12, vainstein_pa14}, as were various coevolutionary models \cite{perc_bs10}, involving network topology, noise, and aspiration \cite{pacheco_prl06, pacheco_jtb06, wu_zx_pre06, pestelacci_lncs08, tanimoto_pa13, biely_pd07, cardillo_njp10, dai_ql_cpl10, hetzer_pone13}, to name but a few examples. In particular, it was found that heterogeneities in the system, sometimes also referred to as diversity \cite{santos_jtb12}, independent of its origin, can significantly enhance cooperation levels in social dilemmas \cite{perc_pre08, perc_pone11, fort_pa08, yao_pha14, szolnoki_epjb08, zhu_p_pone14, szolnoki_pre09d, santos_n08, iwa_pha15}.

A key assumption behind the vast majority of existing research has been, however, that individuals play the same type of game with their neighbors during each interaction. Hashimoto \cite{hashimoto_jtb06, hashimoto_jtb14} was among the first to study so-called multigames, or mixed games \cite{wardil_csf13} (for earlier conceptually related work see \cite{cressman_igtr00}), where different players in the population might adopt different payoff matrices at different times. Considering how  difficult it is to quantify someone's perception of an interaction, it is reasonable to assume that payoff values have numerical fluctuations. Moreover, there is no evidence that the perceived payoff of individuals never changes during their lifetime \cite{wang_z_pre14b}. Based on this, it is natural to analyze games where the payoff matrices are composed of mixtures of different games at different times, as representative of the natural environment where each individual is subject to diverse stimuli. This kind of analysis of multi and mixed games could represent a new line of research in evolutionary game theory, considering the merging of various different games as statistical fluctuations. A complementary approach to the study of mixed games is the study of games on interdependent networks \cite{boccaletti_pr14, santos_md_srep14, baokui_epl14}, where two distinct structured  populations interact via dependency links using different games. A canonical example of a mixed game perspective is how the owner of a cheap car can have a very different risk perception on a highway crossing compared to the owner of a new expensive car. Recent research has revealed that this is an important consideration, which can have far-reaching consequences for the outcome of evolutionary games \cite{wang_z_pre14b, szolnoki_epl14b, amaral_jpa15}.

Here we wish to extend the scope of mixed games \cite{wardil_csf13, amaral_jpa15}, by studying a model where during each interaction individuals play a game that is drawn uniformly at random from an ensemble. In particular, we consider a setup with two different payoff matrices ($G_1$ or $G_2$), and we study evolutionary outcomes on the square lattice, on scale-free and on random networks. As we will show, our results strongly support preceding research that highlights the importance of heterogeneity, as well as the importance of the population structure in ensuring favorable resolutions of social dilemmas. First, however, we proceed with a more detailed description of the studied evolutionary setup.

\section{Mixed games in structured populations}
\label{Model}

\begin{figure}
\centerline{\epsfig{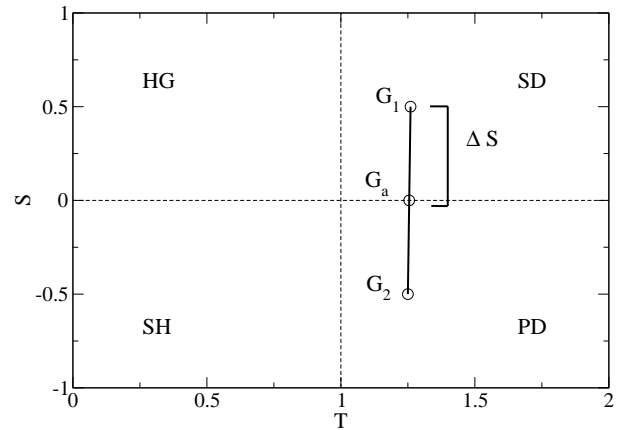}}
\caption{Schematic presentation of the $T-S$ parameter space, as obtained if using $R=1$ and $P=0$. The four evolutionary games are marked in their corresponding quadrants (see main text for details). Moreover, we depict graphically that for each pair of games (SD=$G_1$ and PD=$G_2$, for example), there will be an average game that lies in the middle of the two. Here $\Delta S$ denotes how different $G_1$ is from $G_2$ regarding the parameter $S$.}
\label{fig_parameterspace}
\end{figure}

In the mixed game model, individuals play different games during each interaction. The available strategies are cooperation ($C$) and defection ($D$). The games are represented by the payoff matrix
\[
 \bordermatrix{~ & C & D \cr
                  C & 1 & S \cr
                  D & T & 0 \cr},\]

where $T\in[0,2]$ and $S\in[-1,1]$. The parametrization $G=(T,S)$ spans four different classes of games, namely the prisoner's dilemma game (PD), the snowdrift game (SD), the stag-hunt game (SH), and the harmony Game (HG), as shown in Fig. \ref{fig_parameterspace}.

The mixed game, $G_m$, is defined by the random mixture of two games: $G_1=(T,S_1)$ and $G_2=(T,S_2)$. Each pair of games, $G_1$ and $G_2$, have an average game, $G_a$, given simply by $G_a=(G_1+G_2)/2$. Thus, $T_a=T$ and  $S_a=(S_1+S_2)/2$. Each average game, $G_a$, can be formed by any combination of two games that are symmetrically distributed around it. Different mixtures that correspond to the same average single game can be characterized by the distance $\Delta S$, and each mixed game is from the average single game, namely $\Delta S=|S_1-S_a|=|S_2-S_a|$. We consider $\Delta S$ as a measure of the payoff heterogeneity of the mixed game. In particular, $\Delta S$ defines how far apart $G_1$ is from $G_2$ in the $T-S$ parameter plane (as they are symmetric with respect to $G_a$). Figure~\ref{fig_parameterspace} illustrates this definition schematically. We have chosen to focus on the combination of the prisoner's dilemma and the snowdrift game because they are the most demanding social dilemmas, and also because these two evolutionary games have been studied most commonly in the past. We have verified that our main results remain valid also for other combinations of games on the $T-S$ parameter plane.

We used the Monte Carlo simulation procedure to obtain the dynamics of cooperation in structured populations \cite{szabo_pr07}. The initial configuration is homogeneous: half of the population is $C$ and half is $D$, distributed uniformly at random. For a Monte Carlo step (MCS), each player collects the payoff from all of its direct neighbors. In each pairwise interaction, we randomly choose the matrix $G_1$ with probability $w$, or the matrix $G_2$ with probability  $1-w$, to be the game that is played between the two players during this particular interaction. After the payoff of every site is obtained, we assume a copy mechanism that allows sites to change their strategy. The selected site, $i$, randomly chooses one of its neighbors $j$, and copies the strategy of $j$ with probability $p(\Delta u_{ij})$. The probability of imitation is given by the Fermi-Dirac distribution \cite{szabo_pr07}:
\begin{equation}\label{fermi}
p(\Delta u_{ij})=\frac{1}{ 1+e^{-(u_{j}-u_{i})/K} } ,
\end{equation}
where $u_i$ is the payoff of player $i$ and $K$ can be interpreted as the irrationality of the players, which was taken as $0.3$. We did extensive simulations varying the value of $K$, and we have found that all our main results remain qualitatively the same.

We studied two update rules. In the \textit{synchronous update}, one MCS consists of the copy phase applied to every player at the same time. In the \textit{asynchronous update}, one MCS is the repetition, $N$  times ($N$ is  the population size), of the process of randomly choosing a player to copy one of its neighbors. We stress that biological and human processes are usually best described by the asynchronous update \cite{frean_prsb94}. Nevertheless, here we present the results of the synchronous model as a comparison to some properties of the mixed games.

We run the Monte Carlo dynamics until the network achieves a stable state, where the variables fluctuate around a mean value. We average each quantity over many MCS after the stable state is reached, and then repeat the process for many independent samples \cite{huberman_pnas93}.

The population is structured in complex networks and square lattices. The complex networks are generated with the Krapivsky-Redner algorithm \cite{krapivsky_pre01}, a type of growing network with redirection (GNR) method. We initially create a closed loop with 6 vertices, each one having two directed connections. Then we add a new vertex by randomly connecting it to any of the vertices from the network (growing) and then redirect the connection to the ancestor of this vertex with probability $r$ (Redirection). We repeat this process until the network achieves its final size $N$. Using $r=0.5$, we can create a final distribution that has the properties of a scale-free network \cite{albert_rmp02, barabasi_s99}, with average connectivity degree of $2.7$. The Krapivsky-Redner algorithm is useful because it is relatively fast in computational terms and we can easily change $r$ to obtain a random network ($r=0$).

\begin{figure}
\centerline{\epsfig{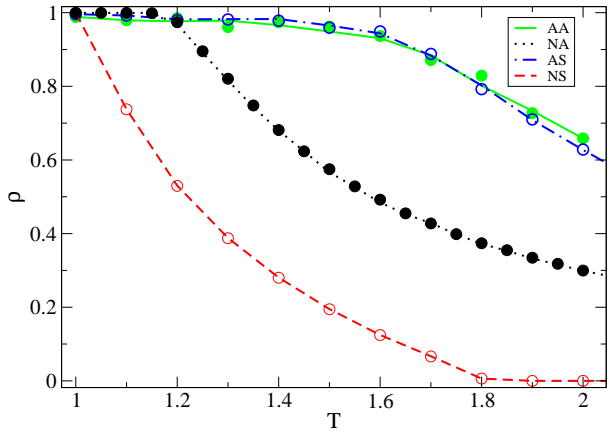}}
\caption{Fraction of cooperation as a function of $T$ for small game heterogeneity. The graph shows simulation results of the mixed game model (symbols) and  the single game defined by the average game (lines) for normalized asynchronous (NA), normalized synchronous (NS), absolute asynchronous (AA) and absolute synchronous (AS). Note that the mixed game has the same behavior as the average game in each model, as long as the game heterogeneity remains small (here $\Delta S <0.1$). The parameters for AA, AS and NA are $S_1=0.1$, $S_2=-0.1$ and $S_a=0$, respectively. The NS model uses $S_1=0.5$, $S_2=0.3$ and $S_a=0.4$.}
\label{fig_resmix}
\end{figure}

In complex networks each player can have a different number of neighbors, which gives rise to a ``topological heterogeneity'' \cite{ohtsuki_jtb06, santos_jeb06, wardil_jpa11}. To analyze how the mixed games are affected by topological heterogeneity, we used two different payoff models, namely the absolute and the normalized value \cite{masuda_prsb07, szolnoki_pa08}. In the absolute value, the total payoff of each player $i$ is just the sum of the payoffs obtained in the interactions with the direct neighbors, denoted as $\{\Omega i\}$:
\begin{equation}\label{payof}
u_i=\sum_{j \in \{\Omega i\}} G(s_i,s_j),
\end{equation}
where $s_i \in\{C,D\}$ is the strategy of player $i$ and  $G(s_i,s_j)$ is the payoff of player $i$ when strategies $s_i$ and $s_j$ are adopted.
In the normalized payoff, the total payoff of  player $i$ is divided by the number its neighbors:
\begin{equation}\label{payof2}
u_i=\frac{\sum_{j\in \{\Omega i\}} G(s_i,s_j)}{k_i},
\end{equation}
where $k_i$ is number of direct neighbors of player $i$.
The normalized payoff model works on the assumption that maintaining many connections is costly, so the payoff is reduced as you get more neighbors \cite{jackson_jet96}. It is important to notice that, in the absolute payoff model, sites with many connections can achieve total payoffs much greater than the average network payoff. By using this four models (synchronous and asynchronous update rules with either absolute or normalized payoffs), applied to different interaction networks (square lattice, random, and scale-free) we are able to confirm the robustness of our results in a broad range of settings. While we do observe quantitative variations in different setups, qualitatively we always obtain the same results, which are thus robust to differences in the accumulation of payoffs, the updating protocol, and the interaction networks.

\begin{figure}
\centerline{\epsfig{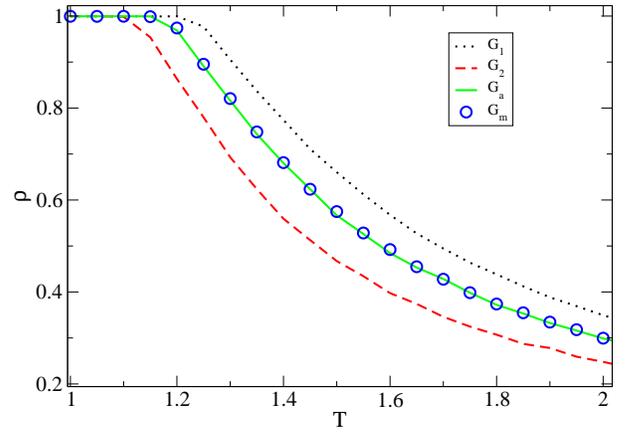}}
\caption{Fraction of cooperation as a function of $T$ for small game heterogeneity in the  normalized asynchronous model. The mixed game is composed of two prisoner's dilemma games. The lines represent single games with parameters $G_1=(T,S_1)=(T,-0.1)$ (doted, black line), $G_2=(T,S_2)=(T,-0.5)$ (dashed, red line), and the average single game $G_a=(T,-0.3)$ (continuous, green line). The mixed game, $G_m$, composed of $S_1$ and $S_2$ is represented by the blue circle. Note that the mixed game $G_m$ behaves as the average game $G_a$.}
\label{fig_mixgame}
\end{figure}

\section{Results}

We analyzed the mixed game model in populations structured in scale-free networks ($r=0.5$), random networks ($r=0.0$), and square-lattices. The population size is with $N=10^{4}$. We used both synchronous and asynchronous Monte Carlo update rules with absolute and normalized payoffs. The MC dynamics was run until the system reaches an equilibrium region where the fraction of cooperation fluctuates around a mean value. The mean value was calculated over  $3\times10^3$ MCS  in the equilibrium region. The transient time needed to reach  equilibrium varies: $7\times10^{3}$ MCS for the normalized asynchronous and the normalized synchronous; $6\times10^{4}$ MCS for the absolute asynchronous; and $4.5\times10^{4}$ MCS for the absolute synchronous. The equilibrium average was then averaged over 100 different networks generated with the same parameters. Note  that the absolute payoff models have a very long relaxation time, compared to the normalized ones. This happens because hubs can obtain huge total payoff's, even when the system is far away from the equilibrium, generating meta-stable states.

\subsection{Small game heterogeneity}
We found that for all  synchronization and payoff rules studied here the final fraction of  cooperators in the mixed game is the same as in the average game as long as the mixed game does not differ much from the average single game, more specifically, as long as the  $\Delta S<0.2$.
Figure~\ref{fig_resmix}  shows all four models used in scale-free networks. The average game is represented by the lines and the mixed game by symbols. Here we used a mixture of Prisoners Dilemma and snowdrift ($S_1=0.1$ and $S_2=-0.1$) for the normalized asynchronous, absolute asynchronous and absolute synchronous models. The normalized synchronous used a combination of two snowdrift games ($S_1=0.5$ and $S_2=0.3$). As can be seem in the figure, the mixed game behaves as the average game for small $\Delta$ in the four models. Figure~\ref{fig_mixgame} shows the asynchronous model in detail, the lines represent the single games $G_1$, $G_2$ and their average $G_a$, while the symbols are for the mixed game ($G_m$) composed by $G_1$ and $G_2$.

\begin{figure}
\centerline{\epsfig{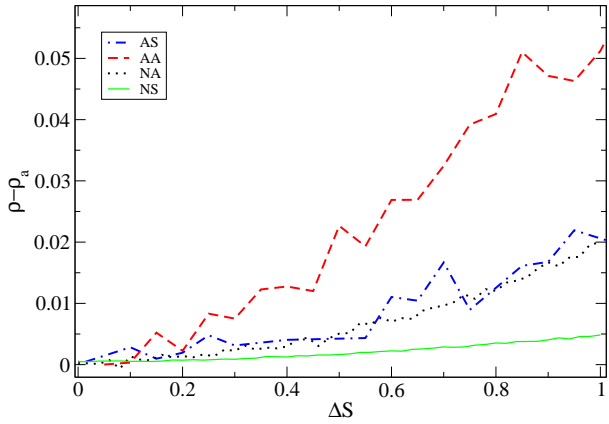}}
\caption{Cooperation increase in mixed games ($\rho$), compared to the average game ($\rho_a$), as game heterogeneity increases in scale-free networks. The average game is the weak prisoner's dilemma. Greater heterogeneity benefits cooperation, although asynchronous models seems to obtain a greater advantage. The weak prisoner's dilemma parameters are $S_a=0$ and $T=1.7$.}
\label{fig_het}
\end{figure}

It is important to notice that each model exhibits different behaviors: different cooperation levels and different critical values of T for the extinction of cooperation (an extensive review can be found in \cite{szabo_pr07}). We do not wish to analyze these differences, as they are well know in the literature. Instead, our goal is to analyze the effect of game heterogeneity in all models.

The average payoff is the same in the average single game and in the mixed game. As in \cite{szolnoki_epl14b, wang_z_pre14b}, the addition of cooperative games together with more selfish games do not change the mean payoff. In contrast, punishment mechanisms often increases cooperation while lowering the average payoff \cite{sigmund_tee07}. The equivalence between the mixed game and the average-single game also holds for different values of $w$. We also studied random networks ($r=0$) and super-hubs networks ($r=1.0$; every vertex is connected to one of the six initial nodes)  \cite{krapivsky_pre01}. For our models the main result shown in Fig.~\ref{fig_mixgame} still holds: the mixed game is equivalent to the average single game in the terms of the final number of cooperators and the average payoff, as long as the game heterogeneity $\Delta S$ is small. This result reinforces what was already know for mixed games in well-mixed populations, rings and square lattices \cite{wardil_csf13, amaral_jpa15}. We point out that the dynamic of cooperation is highly dependent on the topology \cite{ szabo_pr07, santos_prsb06, nowak_aam90, szolnoki_pa08, albert_rmp02, santos_jeb06, santos_prl05, szabo_pre05, tang_epjb06}, nevertheless the mixed game still behaves as the average game. It is very interesting to notice that the topology, irrationality, update rule and copy mechanism drastically alters the final fraction of cooperators, but it seems not to change the equivalence between the average and the mixed game if heterogeneity is small. We found that the only thing that considerably changes this behavior, is how distant the parameters ($S$ or $T$) are from their mean value. In the next section we proceed to study the effect of large game heterogeneity.

\subsection{Large game heterogeneity}

\begin{figure}
\centerline{\epsfig{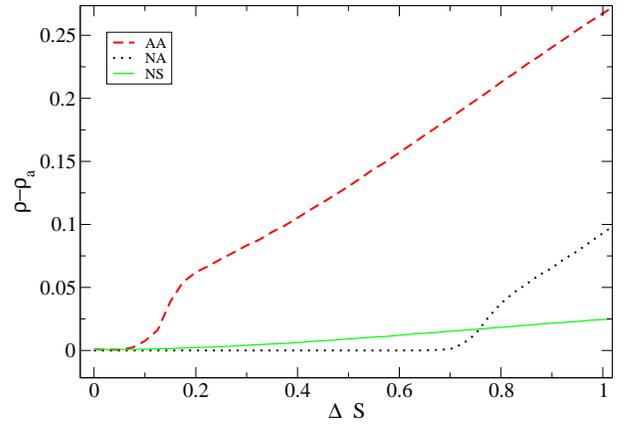}}
\caption{Cooperation increase in mixed games ($\rho$), compared to the average game ($\rho_a$), as game heterogeneity increases in random networks. The behavior of each model is different from the complex networks, Nevertheless the cooperation still benefits from the heterogeneity. We used the weak prisoner's dilemma for the average game and $T=1.8$.}
\label{fig_deltarandom}
\end{figure}

Game heterogeneity in the mixed game model can be measured by $\Delta S$. If the mixed game was equivalent to the average single game for any condition, the variation of $\Delta S$ would be irrelevant. We found impressive results showing that game heterogeneity enhances cooperation, as shown in Fig.~\ref{fig_het} for scale-free networks (in this case, the average game is a weak prisoner's dilemma given by  $S_a=0$ and $T_a=1.7$). Note that although the average game remains the same, the games $G_1$ and $G_2$  become more distinct as $\Delta S$ increases. Obviously, for $\Delta S=0$ we get the trivial case $G_1=G_2=G_a$. It is interesting to notice that an increase in $\Delta S$ favours cooperation despite the fact that, at the same time $G_1$ becomes more ``cooperative'', $G_2$ becomes more ``selfish''. Figure~\ref{fig_deltarandom} shows the effect of large game heterogeneity in random networks for a mixture of games in which the average game is a weak prisoner's dilemma with $T=1.8$. The effect of heterogeneity is stronger in random networks than in scale-free networks. Results from Fig.~\ref{fig_het} and Fig.~\ref{fig_deltarandom} shows that although the final fraction of cooperation can be increased by game heterogeneity, it is highly sensitive to the model.

\begin{figure}
\centerline{\epsfig{file=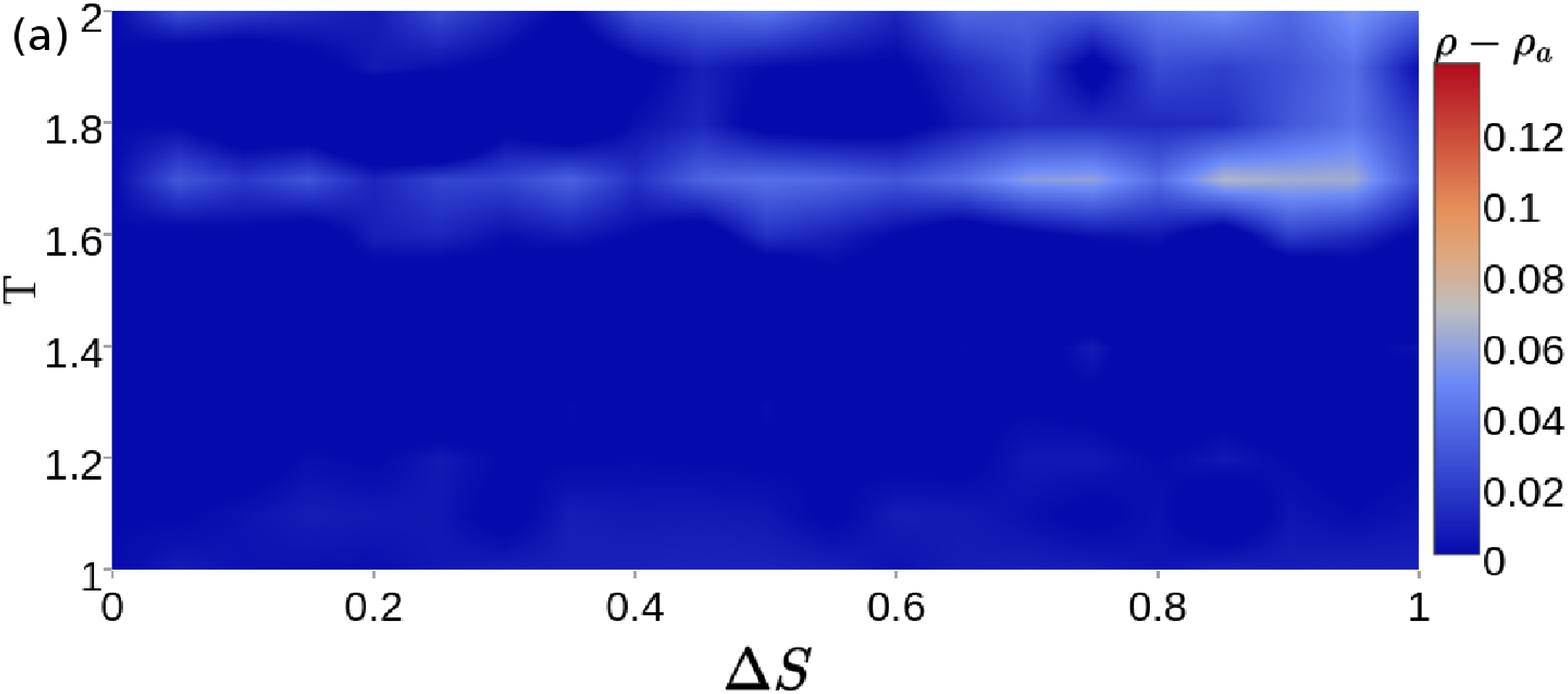,width=8cm}}
\centerline{\epsfig{file=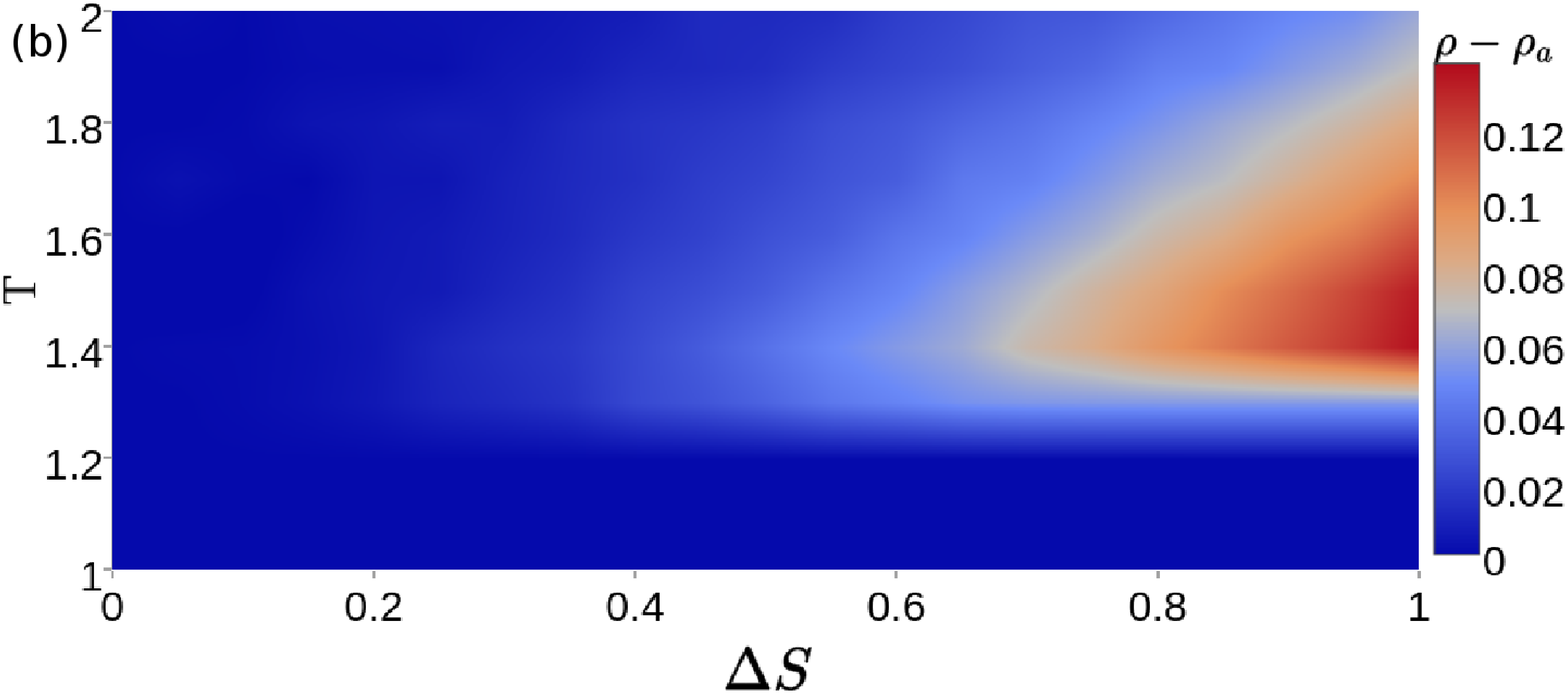,width=8cm}}
\caption{Color map showing how game heterogeneity ($\Delta S$) affects cooperation for various T values in the absolute asynchronous (a) and normalized asynchronous (b) scale-free network model. Each model has an optimum $T$ value, where the evolution of cooperation is enhanced the most.}
\label{fig_SFc}
\end{figure}

\begin{figure}
\centerline{\epsfig{file=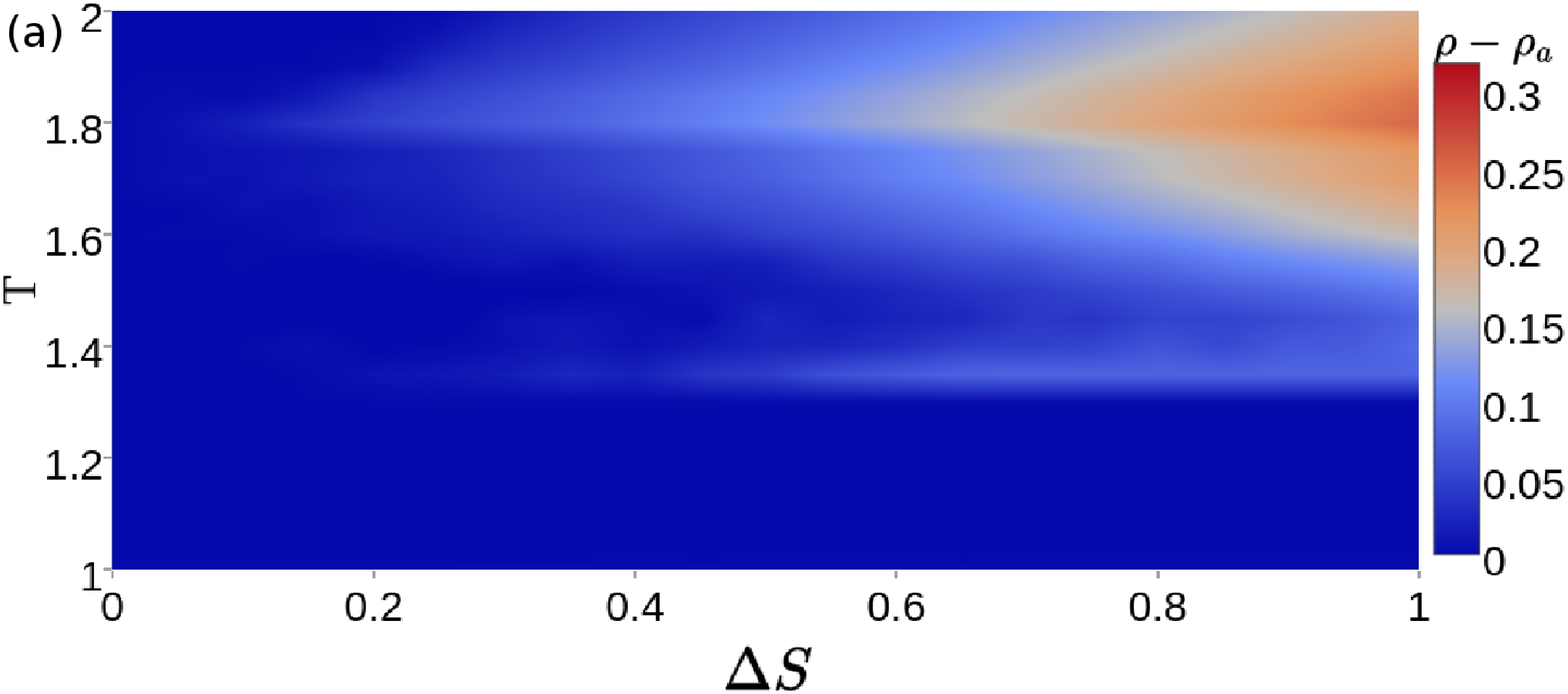,width=8cm}}
\centerline{\epsfig{file=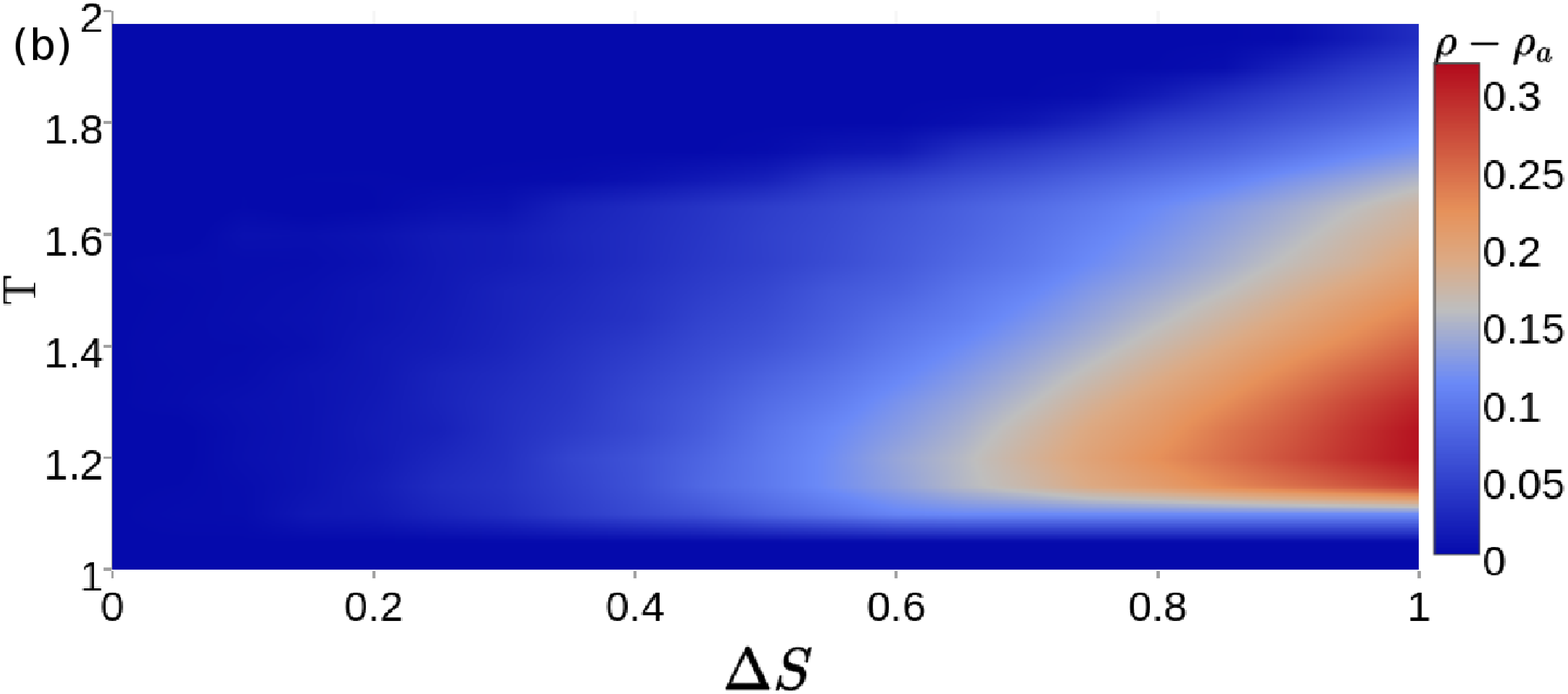,width=8cm}}
\caption{Color map showing how $\Delta S$ affects cooperation in the absolute asynchronous (a) and normalized asynchronous (b) model for different values of $T$ in the random network. It can be seem here,comparing to the scale-free color map, how the optimum $T$ value for increasing cooperation is highly dependent on the topology and synchronization of the model. On this points cooperation can be enhanced in even $0.3$ compared to the average game.}
\label{fig_Randomc}
\end{figure}

The enhancement of cooperation due to game heterogeneity on $S$ happens for all values of $T$, as shown in Fig. \ref{fig_SFc} for scale-free networks.  The increase in $\Delta S$ benefits cooperation, but there are optimal values of T where cooperation is most promoted. For the normalized asynchronous model, a boost of $0.15$ is obtained at $T=1.45$. Figure~\ref{fig_Randomc} shows the fraction of cooperation as a function of $T$ in random networks. The cooperation boost is of almost $0.3$ in some points.

\begin{figure}[b]
\centerline{\epsfig{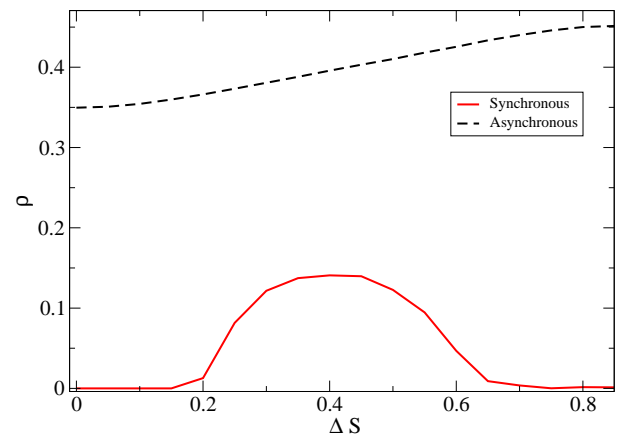}}
\caption{Total fraction of cooperation as $\Delta S$ increases on the square lattice using $T=1.04$. The asynchronous model exhibit enhancements due to game heterogeneity, but for $T=1.04$ the synchronous model should have the cooperation extinct in the average game. Nevertheless, for some $\Delta S$ values the cooperation re-appears even after the extinction threshold.}
\label{fig_deltasq}
\end{figure}

The analysis of the mixed game model in square lattices was very surprising. For $T$ values where cooperation usually survives in the average single game, we found that large game heterogeneity in $S$ promotes cooperation in the mixed game. More interestingly, we found that for the synchronous update cooperation can spontaneously re-emerge even after the critical value of cooperation extinction ($T_c\approx1.04$ for the synchronous model in single games \cite{szabo_pr07}), as shown in Fig.  \ref{fig_deltasq}. Game heterogeneity make the cooperation re-emerge for some values in the range $0.2 < \Delta S < 0.6$. In scale-free and random networks, heterogeneity merely enhanced the fraction of cooperation. But in square lattices, the mixed game is totally dominated by defectors until the heterogeneity reaches $0.2$, when the cooperators re-emerge.

\begin{figure}
\centerline{\epsfig{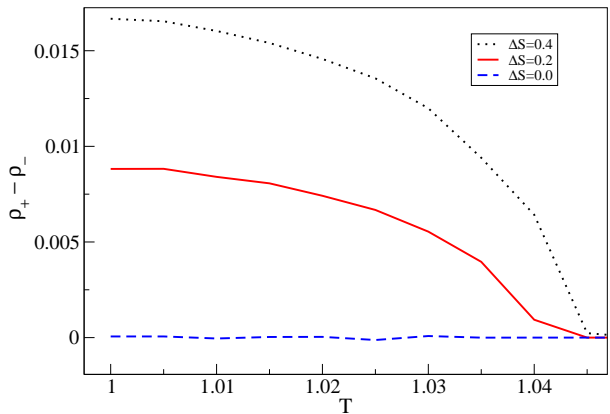}}
\caption{Difference in the populations of cooperators playing the snowdrift ($\rho_+$ ) and the prisoner's dilemma ($\rho_-$) game as $T$ increases for various values of $\Delta S$. Notice the asymmetry, where cooperators play the SD game more frequently. The difference is always positive and grows with increasing values of $\Delta S$.}
\label{fig_transiinfo}
\end{figure}

To understand how game heterogeneity promotes cooperation, we investigate asymmetries introduced by the mixture of PD and SD games on the square lattice. We analyze the histogram describing the number of times each cooperator plays $G_1$ (SD game), or $G_2$ (PD game), during a typical Monte Carlo step. The average game is the weak prisoner's dilemma, $G_1$ favors cooperation ($S>0$), while $G_2$ favors defection ($S<0$). This creates a natural separation in the population\\

\noindent population between SD ($\rho_+$) and PD ($\rho_-$) players. However, the separation is not fixed but changes during the evolution. By recording who plays PD and who plays the SD game, and when, we find that most players usually play both games with equal probability over time. Nevertheless, there are some differences across the population. Results presented in Fig.~\ref{fig_transiinfo} reveal that the fraction of cooperators who play SD more often is higher than the fraction of cooperators who play PD more often. This asymmetry indicates that even if the games are randomly chosen at each step, there is a flux of cooperation towards sites where cooperation is favored. In contrast, defectors do not benefit from the PD population.

The histogram analysis indicates that cooperation enhancement is due to the intrinsic asymmetry between cooperators and defectors. Players that often play the SD game have a higher chance of becoming cooperators, even in the presence of defectors, because of the high positive value of S. Conversely, for greater negative values of S, players that often play the PD game have no incentive to become defectors when they are surrounded by defectors (recall that the $P$ value remains the same). In the long run, players will play PD and SD with the same frequency on average. But, locally, some players can play SD more frequently, increasing their chance to start a cooperation island. The asymmetric effect of the negative S on PD will not cause the opposite, i.e., high negative values of S does not lead to the formation of defectors clusters \cite{wang_z_pre14b,szolnoki_epl14b, yao_pha14}. In time, these small islands of cooperation can grow and eventually become stable, enhancing the cooperation of the model. This is the same asymmetric effect that is observed in heterogeneous multigames \cite{wang_z_pre14b, szolnoki_epl14b, yao_pha14}.

\section{Discussion}
We have studied mixed games on random and scale-free networks, and on the square lattice, focusing specifically on the effect of game heterogeneity. We showed that for small heterogeneity, mixtures of randomly choose games games behave as the average single games, which agrees with previous work using mean-field analysis for the square lattice and ring topologies \cite{wardil_csf13, amaral_jpa15}. We showed that the equivalence between the mixed game and the average single game is still valid for all various topologies, different synchronization rules and different values of irrationality. Nevertheless, our main result is in large game heterogeneity regime, where heterogeneity breaks the equivalence between the mixed game and the average single game and  enhances  cooperation. In particular, the enhancement is highly sensitive to the topology and the applied updating rule used to simulate the evolutionary dynamics. On the square lattice, for example, sufficiently strong heterogeneity resurrects cooperation after the single game extinction threshold value of $T$.

Interestingly, the mean-field model predicts that the mixed game should always behave as the average game \cite{amaral_jpa15}, in contrast to what was found for the networks and the square lattice here. This further highlights the importance of population structure and how cooperation can thrive by simple mechanisms such as network reciprocity and heterogeneity. Future work could study mixed games using a normal distribution of games, instead of just two games. Based on the results presented in this paper, it seems reasonable to expect that the variance of the distribution would affect the final frequency of cooperators.

Finally, we note that our results strongly supports preceding works on how different types of heterogeneity, regardless of origin, promote cooperation \cite{santos_jtb12, perc_pre08, perc_pone11, fort_pa08, yao_pha14, szolnoki_epjb08,zhu_p_pone14, szolnoki_pre09d, santos_n08,lei_c_pa10, sun_l_ijmpc13, tanimoto_pre13, yuan_wj_pone14, iwa_pha15}. While game heterogeneity does offer advantages to both cooperators and defectors, only the former can reap long-term benefits. Defectors are unable to do so because of a negative feedback loop that emerges as their neighbors become weak due to the exploitation. As we have shown, this holds true also for mixed games, and we hope that this paper will motivate further research along this line.

\begin{acknowledgments}
This research was supported by the Brazilian Research Agencies CAPES- PDSE (Proc. BEX 7304/15-3), CNPq and FAPEMIG, and by the Slovenian Research Agency (Grants J1-7009 and P5-0027).
\end{acknowledgments}

\end{document}